%% file: haven_paper.tex
\documentclass[conference]{IEEEtran}
\IEEEoverridecommandlockouts

\usepackage{cite}
\usepackage{amsmath,amssymb,amsfonts}
\usepackage{graphicx}
\usepackage{textcomp}
\usepackage{multirow}
\usepackage{booktabs}
\usepackage{xcolor}
\usepackage{url}
\usepackage[hidelinks]{hyperref}
\usepackage{tikz}
\usetikzlibrary{positioning,arrows.meta,fit,backgrounds,calc,matrix}

\newcommand{\parlabel}[1]{{\smallskip\noindent\textbf{#1}}}

\begin{document}

\title{HAVEN: Hybrid Automated Verification ENgine for UVM Testbench Synthesis with LLMs}

\author{%
\IEEEauthorblockN{Chang-Chih Meng, Yu-Ren Lu, Guan-Yu Lin, Tsung Tai Yeh, Kai-Chiang Wu, I-Chen Wu}
\IEEEauthorblockA{National Yang Ming Chiao Tung University, Hsinchu, Taiwan\\
Email: \{mcc.cs11, yrlu.cs14, ericlin.cs13\}@nycu.edu.tw,
\{ttyeh, kcw, icwu\}@cs.nycu.edu.tw}%
}

\maketitle

\input{sections/abstract}
\input{sections/1_introduction}

\input{sections/2_overview}

\input{sections/3_testbench_generation}

\input{sections/4_sequence_optimization}
\input{sections/5_setup}
\input{sections/6_results}
\input{sections/7_discussion}

\bibliographystyle{IEEEtran}
\bibliography{references}

\end{document}

%% file: sections/abstract.tex
\begin{abstract}


Integrated Circuit (IC) verification consumes nearly 70\% of the IC development cycle, and recent research has begun leveraging Large Language Models (LLMs) to automatically generate testbenches and reduce verification overhead. However, we found that LLMs have difficulty generating testbenches correctly. Unlike high-level programming languages, Hardware Description Languages (HDLs) are extremely rare in LLMs training data, leading LLMs to produce incorrect code.

To overcome challenges when using LLMs to generate Universal Verification Methodology (UVM) testbenches and sequences, we propose HAVEN (Hybrid Automated Verification ENgine) to prevent LLMs from writing HDL directly. 
For UVM testbench generation, HAVEN utilizes LLM agents to analyze design specifications to produce a structured architectural plan. 
The HAVEN Template Engine then combines with predefined and protocol-specific templates to generate all UVM components with correct bus-handshake timing. For UVM sequence generation, HAVEN introduces a Protocol-Aware Sequence Domain-Specific Language (DSL) that decomposes sequences into fine-grained step types. 
A set of predefined DSL patterns first establishes sequences that achieve a high coverage rate without LLM involvement. HAVEN continues to improve the coverage rate by iteratively leveraging LLM agents to analyze coverage gap reports and compose additional targeted DSL sequences.

Unlike previous works, HAVEN is the first system that utilizes pre-defined, protocol-specific Jinja2 templates to generate all UVM components and UVM sequences using our proposed Protocol-Aware DSL and rule-based code generator. 
Our experimental results on 19 open-source IP designs (180--11\,k~LOC) spanning three interface protocols (Direct, Wishbone, AXI4-Lite) show that HAVEN achieves 100\% compilation success, 90.6\% code coverage, and 87.9\% functional coverage on average, and is SOTA among LLM-assisted testbench generation systems.

\end{abstract}

%% file: sections/1_introduction.tex

\begin{figure*}[!t]
\centering
\includegraphics[width=0.95\textwidth]{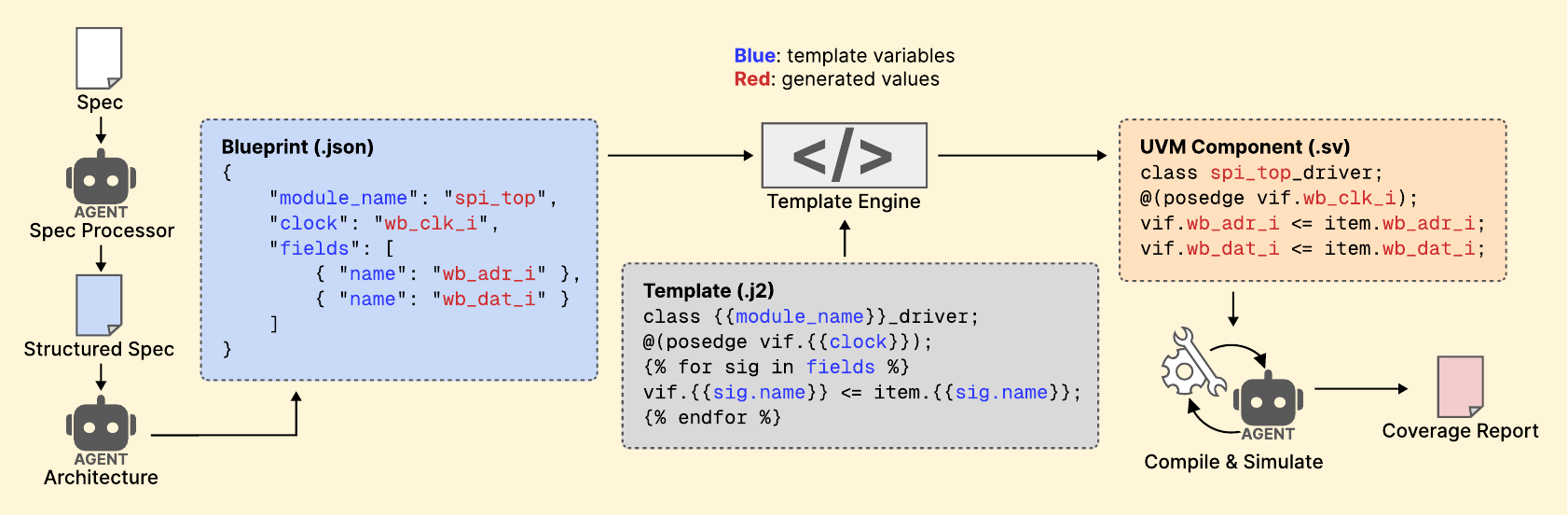}
\caption{Stage~1: Testbench Generation pipeline (driver path shown as a worked example for SPI).
The Spec Processor and Architecture agents convert the design specification into a structured Blueprint (\texttt{.json}).
The Template Engine combines the Blueprint with a pre-defined Jinja2 template (\texttt{.j2}) to produce a protocol-correct UVM component (\texttt{.sv}); the same engine renders other components (monitor, scoreboard, BFM, subscriber) from their respective templates.
A Compile \& Simulate agent then produces the coverage report.}
\label{fig:blueprint}
\end{figure*}

\section{Introduction}
\label{sec:introduction}

Integrated Circuit (IC) Verification, a key process in ensuring that the functional units in IC chips work correctly, consumes nearly 70\% of the IC development cycle~\cite{foster20252024,foster20222022}. 
Conventionally, verification engineers manually develop testbenches and sequences tailored to each IC design's protocol and control logic, thereby incurring substantial verification overhead~\cite{semiconductor2009international}. 
To reduce this overhead, recent research has begun leveraging Large Language Model (LLM) agents to automatically generate testbenches~\cite{zhong2023llm4eda,pan2025survey,xu2024meic,hu2025uvllm,zhang2025llm4dv,thakur2024verigen,liu2023chipnemo,chang2023chipgpt,yan2025assertllm}.

However, LLMs have difficulty generating correct testbenches. Unlike high-level programming languages such as Python, JavaScript, and C++, HDLs are extremely rare in LLM training data. 
For instance, LLMs find it hard to distinguish between blocking (\texttt{=}) and non-blocking (\texttt{<=}) assignments and to employ clock-edge synchronization correctly. 
These constructs have no analog in high-level languages, causing LLMs to frequently produce incorrect code~\cite{xu2025revolution,zhang2026understanding,zhang2025llm,tonmoy2024comprehensive}. 
Recent approaches such as AutoBench, CorrectBench, and ConfiBench~\cite{qiu2024autobench,qiu2025correctbench,qiu2025confibench,tsai2024rtlfixer,yao2025hdldebugger} leverage LLMs to directly write entire testbenches and rely on iterative self-correction to repair errors. 
However, this iterative debugging process is expensive in both time and token usage and may still not yield correct results.

As IC designs grow in complexity, hand-writing monolithic testbenches becomes increasingly difficult to maintain and reuse.
The Universal Verification Methodology (UVM)~\cite{ieee1800ieee}, a widely adopted SystemVerilog-based industry framework, addresses this by decomposing a testbench into standardized, reusable components, such as drivers, monitors, scoreboards, and subscribers. Each UVM component has a well-defined syntax convention. 
When applying LLMs to automate UVM testbench generation, two specific challenges arise.

The first challenge is generating the structural UVM components. We observed that these components adhere to well-defined formats. When the configurations of the Device Under Test (DUT) are altered, we can only update the port list and protocol type in the testbench template. This template-based method is well-suited for generating UVM testbenches.
The prior work, UVM$^2$~\cite{ye2025concept}, employed hand-crafted templates to assist an LLM in generating testbenches. However, the UVM$^2$ template covers only five simple UVM components, without protocol information. 
Consequently, when DUTs contain bus protocols with clocking-sequence events, misusing a blocking assignment in a bus-handshake protocol often causes the entire verification process to fail.

Second, it is also challenging to generate UVM sequences that achieve a high coverage rate. UVM sequences define the test stimuli and their execution ordering to examine the DUT's functional units. Unlike the structural UVM components, sequences are derived from design specifications and vary across DUTs, as each design has its own protocol and control flow. 
Previous work has applied reinforcement learning and Bayesian optimization to search for appropriate test stimuli~\cite{gadde2024efficient,fine2003coverage,kumar2023optimizing,wu2024survey,ioannides2012coverage}.
However, a DUT can contain complex control flows or multi-step bus protocols. For instance, a Wishbone transaction requires a coordinated sequence of address setup, strobe assertion, and acknowledgment waiting.
Although the exploration space grows significantly, the set of stimuli that actually reach uncovered states remains small. These methods, therefore, struggle to converge efficiently, resulting in a reduced IC verification coverage rate.


To address these challenges in automatic UVM testbench generation, we propose \textbf{HAVEN} (Hybrid Automated Verification ENgine)\footnote{HAVEN will be released as open source upon publication; the full implementation, including all protocol templates and DSL CodeGen, will be made publicly available.}, which prevents LLM agents from writing SystemVerilog code directly.
Instead, HAVEN leverages LLMs for what they do best—extracting and organizing information from design specifications—and delegates all code generation to the rule-based code generator (CodeGen) that guarantees SystemVerilog syntactic correctness.

To generate correct UVM components, HAVEN employs LLM agents to analyze design specifications and produce a UVM architectural plan: a structured JSON document that captures the testbench structure (agent topology, component connectivity, interface protocol type) and signal-level data contracts. 
A Template Engine then combines this architectural plan with pre-defined, protocol-specific Jinja2 templates to generate all UVM components, including drivers with correct bus-handshake timing (Section~\ref{sec:pipe_gen}). 
Since the templates encode protocol-correct coding conventions, this approach eliminates SystemVerilog syntax errors when using LLMs to generate UVM components for the testbench. 

To accurately generate UVM sequences and optimize the coverage rate, HAVEN introduces a Protocol-Aware Sequence Domain-Specific Language (DSL) that decomposes sequences into fine-grained step types (e.g., register write, poll, value sweep). 
A rule-based CodeGen translates each DSL step into protocol-correct UVM sequence code, ensuring SystemVerilog syntactic correctness (Section~\ref{sec:pipe_dsl}). 
HAVEN uses the DSL in two stages. 
First, a set of predefined DSL sequence patterns, such as constrained random, field value sweeps, toggle patterns, and FIFO stress tests, is triggered by signal characteristics and protocol structure, without any LLM involvement.
These predefined sequences alone have achieved a sufficient coverage rate.
Second, to minimize remaining coverage gaps, HAVEN employs LLM agents to analyze coverage gap reports and compose additional, targeted DSL sequences (Section~\ref{sec:pipe_seq}).
This process can iterate: after each simulation, HAVEN feeds the updated coverage gaps back to the LLM, which generates new DSL sequences targeting uncovered behaviors, until coverage converges or a maximum iteration count is reached.



We evaluate HAVEN on 19 open-source IP designs spanning three interface protocols (Section~\ref{subsec:benchmark}).
Our experiments show the superiority of HAVEN in LLM-assisted testbench generation, specifically achieving the following:
\begin{enumerate}
\item \textbf{High compile reliability:} Our template-generated UVM components
      achieve \textbf{100\% compilation success} across all 19 IP designs,
      confirming that rule-based code generation eliminates the syntax
      errors observed in LLM-generated components.
\item \textbf{High coverage:} The DSL sequences achieve
      \textbf{90.6\% average code coverage} and \textbf{87.9\% average
      functional coverage} through iterative refinement, establishing
      state-of-the-art results among LLM-assisted UVM verification
      approaches.
\item \textbf{Fast and low-cost:} The full pipeline completes in a single
      run without per-design tuning, averaging 6 LLM calls, 68k tokens,
      and \$0.38 per design.
\end{enumerate}

The key insight behind these state-of-the-art results is that LLMs should not write HDL testbenches directly.
LLMs excel at understanding design specifications, extracting structured information, and interpreting coverage gaps, but they struggle with the syntactic and timing conventions of HDL code.
By restricting LLMs to structured information extraction and delegating all code generation to rule-based systems, HAVEN achieves both HDL syntax correctness and high coverage with low costs.

%% file: sections/2_overview.tex

\section{HAVEN: An Overview}
\label{sec:overview}

There are two stages in the HAVEN pipeline: Stage~1 generates all UVM components, including predefined sequences; Stage~2 iteratively improves the coverage rate through LLM-driven DSL sequence generation.

In Stage~1 (Section~\ref{sec:pipe_gen}), the goal is to generate a complete, compile-passing UVM testbench with predefined sequences.
To achieve this, LLM agents extract structural and protocol information from the design specification into two structured JSON artifacts: the UVM Blueprint and Protocol Flows.
A Template Engine then combines these artifacts with pre-defined, protocol-specific templates to generate all UVM components, including drivers with correct bus-handshake timing, peripheral Bus Functional Models (BFMs), and functional coverage subscribers; Fig.~\ref{fig:blueprint} illustrates this with the driver path as a worked example.
Stage~1 also produces predefined sequences through rule-based strategy inference, without any LLM involvement.
A compile-fix loop ensures the testbench compiles correctly before proceeding to Stage~2.

In Stage~2 (Section~\ref{sec:pipe_seq}), the goal is to close the remaining coverage gaps that predefined sequences cannot reach.
To do this, the LLM analyzes the coverage gap report from Stage~1's simulation and expresses targeted sequences in the structured DSL JSON format (Fig.~\ref{fig:stage2_loop}).
A rule-based DSL CodeGen translates the JSON into protocol-correct UVM sequences, which accumulate with Stage~1's predefined sequences.
HAVEN can iterate Stage~2 after each simulation: the updated coverage gap report is fed back to the LLM, which generates additional DSL sequences targeting uncovered behaviors, until coverage converges or a maximum iteration count is reached.

\textbf{Protocol-Aware Sequence DSL}
\label{sec:pipe_dsl}
Both stages generate sequences through the same Protocol-Aware Sequence DSL.
The DSL defines ten step types covering the primary space of verification operations (Table~\ref{tab:dsl_steps}), such as register writes, polling loops, value sweeps, and BFM actions.
To produce a UVM sequence, the DSL source (whether generated by rule-based inference in Stage~1 or by the LLM in Stage~2) is expressed as a JSON document composed of these step types.
A shared rule-based CodeGen then translates the JSON into protocol-correct SystemVerilog~\cite{wang2023grammar,geng2025generating}.
This design ensures that the LLM never writes SystemVerilog directly; it only selects step types and fills in parameters (addresses, values, conditions), while the CodeGen guarantees syntactic and protocol correctness.
Fig.~\ref{fig:stage2_loop} illustrates this translation with a concrete example: an ETHMAC MDIO register read expressed as two DSL steps (\texttt{register\_write} and \texttt{poll}) and the protocol-correct UVM sequence code produced by the CodeGen.
The translation mechanics are detailed in Section~\ref{sec:pipe_seq}.

\begin{table}[t]
\centering
\caption{Protocol-Aware Sequence DSL: step types.}
\label{tab:dsl_steps}
\footnotesize
\setlength{\tabcolsep}{3pt}
\begin{tabular}{l|l}
\hline
\textbf{Step Type} & \textbf{Description} \\
\hline
\texttt{register\_write}  & Write value to register address \\
\texttt{register\_read}   & Read register, store result \\
\texttt{poll}              & Repeated read until condition met \\
\texttt{randomize\_send}  & CRV transaction with inline constraints \\
\texttt{delay}             & Wait $N$ clock cycles \\
\texttt{memory\_write}    & Bulk transfer via BFM backdoor \\
\texttt{bfm\_action}      & Trigger predefined BFM task \\
\texttt{config\_sweep}    & Cartesian product over register fields \\
\texttt{value\_sweep}     & Iterate field through value list \\
\texttt{toggle\_pattern}  & Systematic bit-toggle sequences \\
\hline
\end{tabular}
\end{table}

%% file: sections/3_testbench_generation.tex

\section{Stage~1: Testbench Generation}
\label{sec:pipe_gen}

\begin{figure*}[!t]
\centering
\includegraphics[width=0.8\textwidth]{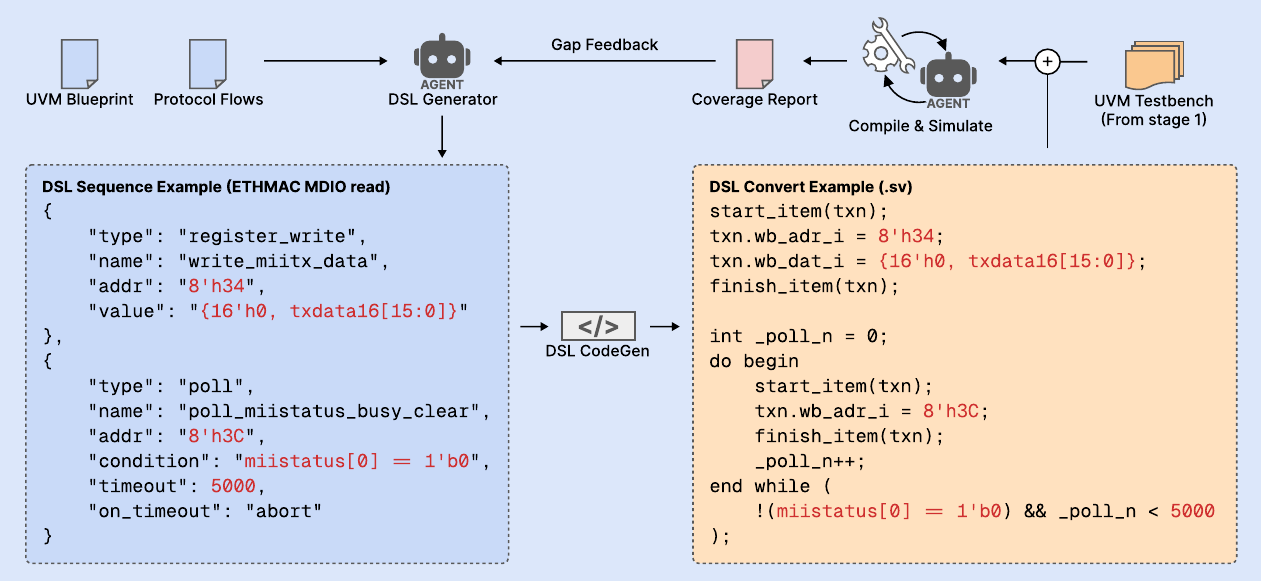}
\caption{Stage~2: Sequence Optimization pipeline.
The DSL Generator agent reads the Structured Spec, UVM Blueprint, Protocol Flows, and the coverage gap report from Stage~1's simulation to produce targeted DSL sequences as JSON.
The rule-based DSL CodeGen (with safety filters) translates them into protocol-correct UVM sequences, which are combined with the Stage~1 testbench and re-simulated by the Compile \& Simulate agent.
The DSL Sequence Example box shows a real ETHMAC MDIO register read (a \texttt{register\_write} followed by a \texttt{poll}); the DSL Convert Example box shows the corresponding generated UVM sequence code.
The Gap Feedback arrow forms the iterative refinement loop.}
\label{fig:stage2_loop}
\end{figure*}

Stage~1 takes the design specification as input and produces a compile-passing UVM testbench with predefined sequences.
The HDL source code of the DUT is not required for testbench generation itself; it is only needed when compiling and simulating the testbench against the DUT.
Fig.~\ref{fig:blueprint} shows the driver path of this pipeline as a worked SPI example: the Spec Processor and Architecture agents extract a Blueprint, the Template Engine renders the corresponding UVM driver from a Jinja2 template, and the Compile \& Simulate agent produces the coverage report.
The full pipeline proceeds through four steps: the LLM extracts structural information into a JSON Blueprint, a rule-based engine infers stimulus strategies, the Template Engine renders all UVM components and predefined sequences, and a compile-fix loop ensures correctness.

\begin{figure}[!t]
\centering
\includegraphics[width=0.4\textwidth]{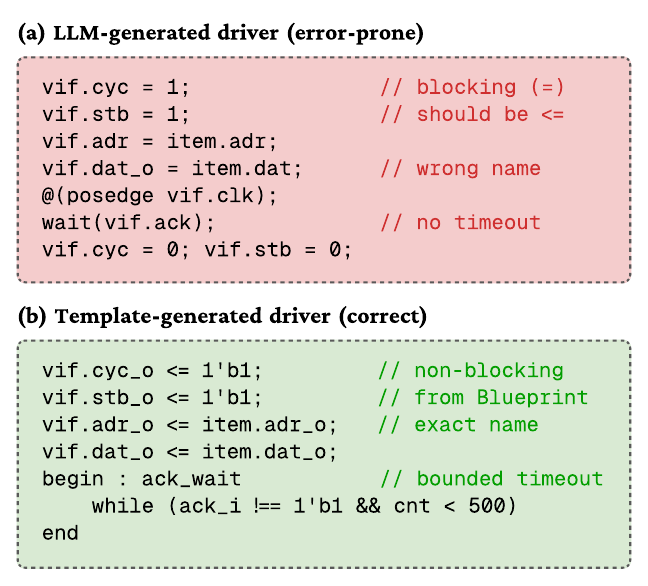}
\caption{Driver comparison.
(a)~LLM-generated code contains three common errors: blocking
assignments, wrong signal names, and an unbounded wait.
(b)~HAVEN's template generates protocol-correct code with
non-blocking assignments, Blueprint-matched signal names, and a
bounded ACK timeout.}
\label{fig:with_without}
\end{figure}

\subsection{Blueprint Extraction}
Before generating any code, the LLM first extracts key information from the design specification and organizes it into a structured JSON document called the UVM Blueprint.
The Blueprint contains agent topology, \texttt{seq\_item} field data contracts (type, width, direction, role), bus protocol type, and Bus Functional Model (BFM) requirements.
Pipeline filters automatically exclude non-stimulus signals (clock, reset, pad, bus-handshake) from \texttt{seq\_item} fields.

\subsection{Strategy Inference}
\label{sec:strategy_inference}
With the Blueprint in hand, HAVEN must decide how to stimulate each signal.
Based on each \texttt{seq\_item} field's role and width as recorded in the Blueprint, a rule-based engine infers the stimulus strategy without further LLM involvement. Fields with width $w \leq 4$ bits receive the \texttt{enumerate} strategy and are swept through all $2^w$ values; wider fields receive the Constrained Random Verification (CRV) strategy; config fields with an explicit default are pinned as \texttt{fixed} (non-rand).

\subsection{Template Rendering and Predefined Sequence Generation}
With the Blueprint and stimulus strategies defined, this step produces all UVM components and predefined sequences.
The Template Engine renders all UVM components from pre-defined Jinja2 templates parameterized by the Blueprint (Fig.~\ref{fig:blueprint}).
Protocol-specific driver templates cover three bus protocols: Wishbone master, AXI4-Lite master, and Direct interface (with ready/done/busy handshake variants).
Fig.~\ref{fig:with_without} contrasts a typical LLM-generated Wishbone driver (with blocking assignments, mismatched signal names, and an unbounded wait) against the template-generated version, which produces protocol-correct code by filling Jinja2 variables (\texttt{\{\{...\}\}}) from the Blueprint.

Stage~1 also produces six types of predefined sequences through the DSL CodeGen (Section~\ref{sec:pipe_dsl}), requiring no LLM involvement (Table~\ref{tab:det_seq}).
The rule-based strategy inference (Section~\ref{sec:strategy_inference}) generates DSL JSON for each sequence type, and the shared CodeGen translates them into protocol-correct SystemVerilog.
Three safety filters are applied during this translation: non-rand constraint removal (constraints on output fields are silently dropped), field validation (references to non-existent signals are rejected), and protocol-timing enforcement (bus-handshake invariants are guaranteed by the driver template, not the sequence).
These predefined sequences reach 84.6\% average code coverage and 79.8\% average functional coverage across 19 IP designs before any LLM-generated sequences are added.

\begin{table}[]
\centering
\caption{Predefined sequences in Stage~1 (no LLM).}
\label{tab:det_seq}
\footnotesize
\setlength{\tabcolsep}{2pt}
\begin{tabular}{l|l|l}
\hline
\textbf{Seq.} & \textbf{Description} & \textbf{Trigger} \\
\hline
CRV     & Random writes, reads, mixed    & Always \\
Enum    & Sweep $2^w$ values ($w\!\leq\!4$) & Small fields \\
Toggle  & Walking-1/0, alternating       & Bus/reg detected \\
FIFO    & Fill, drain, overflow, push/pop & TX/RX reg detected \\
Bank    & Multi-bank seq./interleaved    & Bank addr detected \\
BFM     & Protocol-specific actions      & BFM config present \\
\hline
\end{tabular}
\end{table}

\textbf{Peripheral BFM generation.}
Many DUTs cannot run in isolation: an Ethernet MAC requires an MII PHY model, an SPI master requires a slave device, and a memory controller requires an SDRAM model that responds to bus transactions.
HAVEN generates these peripheral models by extending the Blueprint with a BFM declaration that captures the peripheral type and its connection to the DUT.
The Template Engine renders the corresponding BFM from a library of seven peripheral types (GPIO, I$^2$C slave, MII PHY, SDRAM model, SPI slave, UART serial, and Wishbone slave), and the top-level testbench template wires the BFM to the DUT automatically.
Because BFMs are template-generated, they encode protocol-correct response timing without any LLM-written SystemVerilog.

\textbf{Functional coverage generation.}
Stage~1 also generates functional coverage for every agent.
The subscriber template emits a covergroup for each \texttt{seq\_item} field, with bins derived from the Blueprint's \texttt{cover\_bins} declarations (semantic ranges or width-based automatic binning).
A three-layer signal consistency check ensures that the RTL interface, transaction interface, and monitor mapping all agree on signal names and widths; Blueprints that fail this check are rejected before code generation, eliminating phantom coverpoints.

\subsection{Compile-Fix and Simulation}
Before proceeding to Stage~2, the generated testbench must compile and simulate correctly.
Template-generated UVM components form a protected set that the compile-fix loop never modifies.
When compilation fails, the LLM reads the VCS error log and corrects the LLM-generated sequence item; fixes are resolved in dependency order for up to five iterations.
Once the testbench compiles, HAVEN runs a simulation to produce the initial coverage report, which serves as the starting point for Stage~2's iterative refinement.

%% file: sections/4_sequence_optimization.tex

\begin{table*}[t]
\caption{Benchmark suite, code coverage, functional coverage (\%), and token cost on 19 IP designs. The UVM$^2$ column marks ($\checkmark$) the nine designs that overlap with the UVM$^2$ benchmark~\cite{ye2025concept}. $K$ denotes the number of LLM-driven DSL sequence generation iterations in Stage~2 (each iteration feeds the latest coverage gap report back to the LLM); we use $K{=}3$ throughout.}
\label{tab:coverage}
\centering
\footnotesize
\setlength{\tabcolsep}{3pt}
\begin{tabular}{l|r|r|l||r|r||r|r||c||r|r|r|r}
\hline
 & \multicolumn{3}{c||}{\textbf{Benchmark}} & \multicolumn{2}{c||}{\textbf{Code Cov.\ (\%)}} & \multicolumn{2}{c||}{\textbf{Func.\ Cov.\ (\%)}} & \textbf{UVM$^2$} & \multicolumn{4}{c}{\textbf{Token Cost}} \\
\textbf{Design} & \textbf{LOC} & \textbf{Ports} & \textbf{Sec.\ I/F} & \textbf{Stage~1} & \textbf{+Stage~2} & \textbf{Stage~1} & \textbf{+Stage~2} & \textbf{code/func} & \textbf{Calls} & \textbf{In} & \textbf{Out} & \textbf{USD} \\
\hline\hline
\multicolumn{13}{l}{\emph{Direct Interface}} \\
\hline
ALU         &    482 &  9 & Handshake  & 86.8 & \textbf{96.7} & 55.6 & \textbf{72.2} & $\checkmark$ & 6  & 41{,}384 & 18{,}610 & \$0.33 \\
AES         &    586 & 11 & Handshake  & 96.6 & \textbf{96.6} & 78.6 & \textbf{92.9} & $\checkmark$ & 6  & 39{,}493 & 23{,}010 & \$0.39 \\
SHA3        &    499 &  7 & Streaming  & 94.8 & \textbf{94.8} & 68.3 & \textbf{78.3} & $\checkmark$ & 7  & 37{,}671 & 17{,}661 & \$0.31 \\
SM4         &    420 &  5 & Handshake  & 97.3 & \textbf{97.3} & 100.0 & \textbf{100.0} & $\checkmark$ & 6 & 30{,}949 & 10{,}861 & \$0.21 \\
HUF         &  1{,}572 &  8 & Streaming  & 76.4 & \textbf{76.5} & 80.0  & \textbf{80.0}  & $\checkmark$ & 7 & 34{,}695 & 13{,}980 & \$0.26 \\
DFI         &  1{,}231 & 32 & DDR3 PHY   & 41.6 & \textbf{89.2} & 62.6  & \textbf{96.7}  & $\checkmark$ & 8 & 53{,}769 & 29{,}103 & \$0.50 \\
\hline
\emph{Avg}  &    --- & --- & ---       & 82.3 & \textbf{91.9} & 74.2 & \textbf{86.7} & --- & 7 & 39{,}660 & 18{,}871 & \$0.33 \\
\hline\hline
\multicolumn{13}{l}{\emph{Wishbone Bus}} \\
\hline
SPI         &    352 & 13 & SPI        & 94.2 & \textbf{95.3} & 100.0 & \textbf{100.0} & $\checkmark$ & 6  & 49{,}648 & 19{,}377 & \$0.36 \\
Simple~SPI  &    457 & 13 & SPI        & 92.1 & \textbf{92.1} & 90.0  & \textbf{93.8}  & ---          & 4  & 24{,}086 & 10{,}653 & \$0.19 \\
UART        &    683 & 18 & Serial     & 64.6 & \textbf{88.9} & 100.0 & \textbf{100.0} & $\checkmark$ & 7  & 55{,}977 & 26{,}164 & \$0.46 \\
I2C         &  1{,}281 & 17 & I2C      & 75.5 & \textbf{75.5} & 100.0 & \textbf{100.0} & ---          & 4  & 27{,}079 & 14{,}776 & \$0.25 \\
GPIO        &  1{,}462 & 17 & GPIO     & 94.9 & \textbf{94.9} & 80.0  & \textbf{83.3}  & ---          & 7  & 43{,}021 & 17{,}798 & \$0.32 \\
CAN         &  6{,}593 & 24 & CAN      & 81.9 & \textbf{83.4} & 81.5  & \textbf{81.6}  & ---          & 8  & 83{,}864 & 33{,}455 & \$0.62 \\
ETHMAC      & 11{,}154 & 40 & MII      & 72.8 & \textbf{81.8} & 77.1  & \textbf{77.1}  & ---          & 11 & 69{,}934 & 26{,}956 & \$0.50 \\
SDRAM       &  3{,}929 & 36 & SDRAM PHY & 81.4 & \textbf{84.0} & 54.9  & \textbf{96.4}  & $\checkmark$ & 6  & 66{,}219 & 32{,}208 & \$0.57 \\
\hline
\emph{Avg}  &    --- & --- & ---       & 82.2 & \textbf{87.0} & 85.4 & \textbf{91.5} & --- & 7  & 52{,}479 & 22{,}673 & \$0.41 \\
\hline\hline
\multicolumn{13}{l}{\emph{AXI4-Lite}} \\
\hline
AXIL~RAM    &    180 & 21 & Memory     & 92.1 & \textbf{92.1} & 67.2 & \textbf{80.6} & --- & 6 & 40{,}052 & 25{,}297 & \$0.42 \\
UE~GPIO     &    686 & 23 & GPIO       & 98.0 & \textbf{98.0} & 90.0 & \textbf{92.9} & --- & 7 & 55{,}203 & 20{,}511 & \$0.38 \\
UE~SPI      &  1{,}156 & 33 & SPI      & 78.0 & \textbf{94.2} & 79.6 & \textbf{79.6} & --- & 7 & 57{,}713 & 28{,}746 & \$0.50 \\
UE~Timer    &    542 & 20 & Timer      & 96.2 & \textbf{97.9} & 81.1 & \textbf{94.4} & --- & 6 & 31{,}349 & 11{,}924 & \$0.22 \\
UE~UART     &    752 & 22 & Serial     & 92.7 & \textbf{92.7} & 70.6 & \textbf{70.6} & --- & 4 & 29{,}617 & 15{,}042 & \$0.26 \\
\hline
\emph{Avg}  &    --- & --- & ---       & 91.4 & \textbf{95.0} & 77.7 & \textbf{83.6} & --- & 6 & 42{,}787 & 20{,}304 & \$0.36 \\
\hline\hline
\textbf{Overall (19)} & --- & --- & --- & 84.6 & \textbf{90.6} & 79.8 & \textbf{87.9} & --- & 6 & 45{,}880 & 20{,}849 & \$0.37 \\
\hline
\textbf{Avg (9)}      & --- & --- & --- & --- & \textbf{91.0} & --- & \textbf{90.7} & 87.44 / 89.58$^\dagger$ & --- & --- & --- & --- \\
\hline
\end{tabular}
\vspace{2pt}

{\scriptsize Stage~1 = testbench + predefined sequences. +Stage~2 = after $K{=}3$ rounds of targeted DSL sequence generation, where $K$ is the number of LLM-driven iterations that each feed the updated coverage gap report back to the LLM to produce additional targeted DSL sequences.
Token cost: combined Stage~1 + Stage~2 LLM calls on GPT-5.2 (\$1.75 per 1M input, \$14 per 1M output).
The UVM$^2$ code/func column marks ($\checkmark$) the nine designs that overlap with the UVM$^2$ benchmark; the Avg~(9) row averages HAVEN's coverage on those nine designs, and the UVM$^2$ code/func cell in that row reports UVM$^2$'s own averages on the same nine designs.
$^\dagger$UVM$^2$ averages are reported in the UVM$^2$ paper~\cite{ye2025concept}, not reproduced.} \end{table*}

\section{Stage~2: Sequence Optimization}
\label{sec:pipe_seq}

While Stage~1's predefined sequences already reach 84.6\% average code coverage and 79.8\% average functional coverage on their own, IC designs including complex protocols or deep state machines still require more targeted stimuli that rule-based generation cannot produce to improve their coverage rates~\cite{tasiran2002coverage,naveh2007constraint}.
Stage~2 addresses this gap through a three-step process (Fig.~\ref{fig:stage2_loop}) executed over $K$ iterations, where each iteration feeds the latest coverage gap report back to the LLM; we report $K{=}3$ throughout the evaluation.

\subsection{DSL Sequence Generation}
The goal of the DSL Sequence Generation is to identify what remains uncovered and compose targeted DSL sequences to exercise those paths.
The LLM receives the design specification's protocol flows, the UVM Blueprint, and the coverage gap report from Stage~1's simulation.
The coverage gaps are parsed from the URG output into typed records (line hits, toggle directions, FSM state/transition misses, branch conditions, and functional coverage bins) and presented as a plain-text list so the LLM can process them without dealing with raw EDA tool formatting.
The prompt also includes existing DSL sequences from prior iterations so the LLM can reference what has already been tested and avoid generating duplicates.
The LLM's output is constrained to a JSON schema, ensuring that every response is a valid DSL document; the LLM cannot produce free-form SystemVerilog even if it tries.

To illustrate, consider CAN's Stage~2.
The coverage report lists coverpoint \texttt{FP\_002} (transmission status) as uncovered.
The LLM traces this to a missing TX buffer write followed by a status-register check, and produces a \texttt{transmit\_frame\_via\_tx\_buffer} sequence: a \texttt{register\_write} to load the TX buffer, followed by a \texttt{poll} on status register bit~3 to confirm completion.
The entire output is DSL JSON; the LLM writes no SystemVerilog.

\subsection{Rule-Based Code Generation}
Once the LLM has produced DSL JSON, it must be translated into protocol-correct UVM sequence code.
Before code generation, an auto-fix pass silently repairs common LLM mistakes in the DSL JSON rather than rejecting them, avoiding an extra LLM round-trip.

\begin{figure}[!t]
\centering
\includegraphics[width=0.4\textwidth]{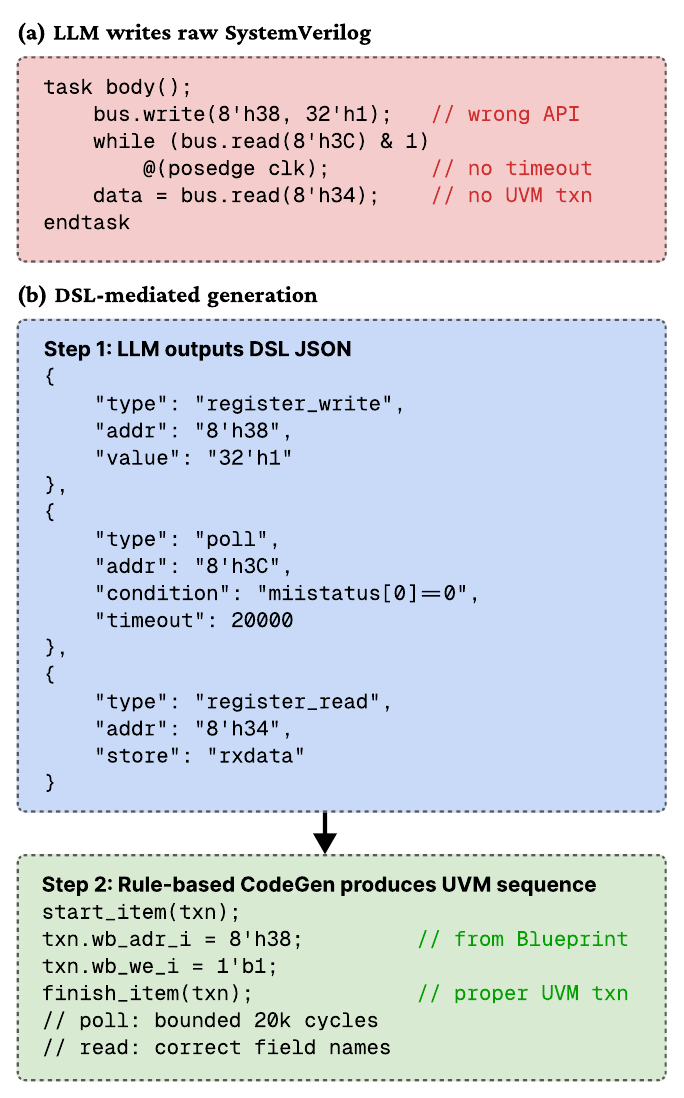}
\caption{Sequence generation comparison.
(a)~LLM-generated SystemVerilog uses wrong API, has no UVM transaction handling, and contains an unbounded wait.
(b)~DSL: the LLM writes only JSON step types; the rule-based CodeGen produces correct UVM transactions with Blueprint-matched signal names and bounded timeouts.}
\label{fig:dsl_vs_raw}
\end{figure}

The shared CodeGen then translates each validated DSL step into protocol-correct UVM code (Fig.~\ref{fig:stage2_loop}).
For a \texttt{register\_write}, it wraps the operation in \texttt{start\_item}/\texttt{finish\_item}, maps the address and value to the correct \texttt{seq\_item} fields by looking up the Blueprint's signal names (e.g., \texttt{wb\_adr\_i}, \texttt{wb\_dat\_i}), and inherits bus-handshake timing from the driver template.
For the \texttt{poll} step, the CodeGen emits a bounded do-while loop that reads the target register until the specified condition is met or the timeout count is reached, preventing the unbounded waits that LLM-generated code often produces.
Because the CodeGen always references the Blueprint for signal names and the driver template for protocol timing, every translated step is guaranteed to be consistent with the rest of the testbench.
DSL step-type analysis across the 19 designs shows that the top five types cover 96.8\% of all LLM-emitted steps, confirming DSL compactness; 3 of the 19 designs (15.8\%) hit the DSL expressiveness ceiling, requiring control-flow primitives beyond the current linear DSL.
Fig.~\ref{fig:dsl_vs_raw} contrasts DSL generation against direct LLM generation.

\subsection{Accumulation and Iterative Refinement}
The generation--simulation loop described in IV-A and IV-B can run multiple times.
We observed that each additional iteration continues to improve coverage, because the updated gap report gives the LLM progressively more focused targets.

The generated UVM sequences accumulate with Stage~1's predefined sequences and are simulated together; new sequences never replace existing ones.
This accumulation design is deliberate: predefined sequences already cover the basic stimulus space (CRV, enumerations, toggles), and each iteration adds incremental coverage on top.
Replacing earlier sequences would risk losing already-covered states.

The same compile-fix loop from Stage~1 ensures the accumulated testbench compiles correctly before each simulation.
After each simulation, the system re-parses the updated coverage gap report and feeds the remaining gaps back to the LLM~\cite{zhang2026llm4cov,nadimi2025tb}, which generates additional DSL sequences targeting uncovered behaviors.
The loop terminates under two conditions: either the coverage improvement between consecutive iterations falls below 0.1 percentage points (early stop), or a maximum iteration count is reached.
In practice, most designs converge within 2--3 iterations.
Per-design coverage results are reported in Section~\ref{sec:results}.

%% file: sections/5_setup.tex

\section{Evaluation}
\label{sec:setup}

We evaluate HAVEN along three dimensions: (1)~compile reliability (does the generated testbench always compile successfully?), (2)~effective coverage (how much code and functional coverage does it achieve?), and (3)~comparison with prior work (how does it compare to UVM$^2$ and other LLM-based approaches?).

\subsection{Benchmark Suite}
\label{subsec:benchmark}

We evaluate HAVEN on 19 open-source IP designs grouped by three bus protocols (Direct, Wishbone, and AXI4-Lite); nine of these designs (ALU, AES, SHA3, SM4, HUF, DFI, SPI, UART, SDRAM) overlap with the UVM$^2$ benchmark~\cite{ye2025concept}.
Per-design size, port count, and secondary interface are listed alongside the coverage results in Table~\ref{tab:coverage}.
Three direct-interface designs use handshake or streaming I/O; eight Wishbone designs add a standardized bus with peripheral-specific secondary interfaces; and five AXI4-Lite designs from the Ultra-Embedded collection exercise the third protocol template.
Design complexity ranges from 180 to 11\,k~LOC, comparable to or exceeding the scale of existing benchmarks~\cite{jin2025realbench,wan2026fixme}.
The Wishbone group intentionally includes two SPI controllers (OpenCores SPI and Simple~SPI) and two GPIO controllers (OpenCores GPIO and UE~GPIO) to test whether the same driver template generalizes across designs of varying complexity.
Designs are sourced from OpenCores\footnote{\url{https://github.com/fabriziotappero/ip-cores}}, Ultra-Embedded\footnote{\url{https://github.com/ultraembedded/core_soc}}, verilog-axi\footnote{\url{https://github.com/alexforencich/verilog-axi}}, and four custom designs (ALU, AES, SPI, UART).

\subsection{Methodology and Metrics}

For each design, we report the result of a single pipeline execution (Stage~1 followed by Stage~2) without per-design tuning or manual intervention.

\parlabel{Code coverage.} The unweighted mean of URG's line, condition, toggle, branch, and FSM state/transition metrics, excluding the group metric.

\parlabel{Functional coverage.} URG's group metric, measured against the covergroups emitted by HAVEN's subscriber template.
The per-field covergroup bins are derived from the Blueprint's \texttt{cover\_bins} declarations (Section~\ref{sec:pipe_gen}), so HAVEN both defines and measures the functional coverage targets without any hand-written golden covergroup.
Code coverage, measured by VCS against the RTL source, provides an independent and objective metric unaffected by covergroup design.

\parlabel{Compile success.} A design is counted as compile-successful if VCS produces a simulation binary without errors after the compile-fix loop.

\subsection{Environment}

\textbf{LLM.}\; All pipeline stages use OpenAI GPT-5.2.
No fine-tuning or domain-specific training is applied.

\textbf{EDA tools.}\; Synopsys VCS~2024.09-SP2 for compilation, simulation, and URG coverage analysis.

%% file: sections/6_results.tex

\begin{table*}[!h]
\caption{Coverage (\%) and average token usage per design across LLMs, combined Stage~1 + Stage~2 ($K{=}3$).}
\label{tab:llm_compare}
\centering
\footnotesize
\setlength{\tabcolsep}{4pt}
\begin{tabular}{l|l|l|c|r|r||r|r|r|l}
\hline
\textbf{Model} & \textbf{Type} & \textbf{Params (active)} & \textbf{Done} & \textbf{Code} & \textbf{Func.} & \textbf{Calls} & \textbf{Input} & \textbf{Output} & \textbf{Cost} \\
\hline
GPT-5.2                       & closed, API & ---           & \textbf{19/19} & \textbf{90.6} & \textbf{87.9} & 6 & 45.9\,k & 20.8\,k & \$0.37 \\
Qwen3.5-27B$^\P$              & open, dense & 27B           & 19/19 & 85.8          & 78.3          & 6 & 33.9\,k & 17.5\,k & free$^\ast$ \\
Qwen3.5-35B-A3B$^\S$          & open, MoE   & 35B (3B)      & 18/19 & 80.8          & 63.9          & 7 & 32.0\,k & 17.5\,k & free$^\ast$ \\
gpt-oss-20b$^\dagger$         & open, MoE   & 21B (3.6B)    & 18/19 & 79.9          & 73.2          & 5 & 30.4\,k & 30.2\,k & free$^\ast$ \\
gpt-oss-120b$^\dagger$        & open, MoE   & 117B (5.1B)   & 17/19 & 81.1          & 68.2          & 4 & 25.7\,k & 16.7\,k & free$^\ast$ \\
Qwen3.5-9B$^{\ddagger\P}$     & open, dense & 9B            & 11/19 & 87.3          & 74.8          & 6 & 34.3\,k & 19.2\,k & free$^\ast$ \\
\hline
\end{tabular}
\vspace{2pt}

{\scriptsize Averages are over 19 IP designs; token usage is per-design average across Stage~1 + Stage~2 LLM calls.
$^\ast$Open-source models are self-hosted via vLLM on an NVIDIA RTX PRO 6000 (96\,GB) and incur no API fee.
$^\dagger$Both gpt-oss models fail on ETHMAC (16\,K output-token limit); gpt-oss-120b additionally fails on SM4.
$^\S$Qwen3.5-35B-A3B fails on DFI.
$^\ddagger$Qwen3.5-9B fails to produce a compiling testbench on 8 of the 19 IP designs; its average is computed over the remaining 11.
$^\P$Qwen3.5 token metrics were added mid-experiment; the Qwen3.5-27B token average is over 13 designs (ALU, CAN, and UE~GPIO runs predate tracking), and the Qwen3.5-9B token average is over the 10 designs that both compiled and were tracked.
Failed designs are excluded from each model's averages.} \end{table*}

\section{Results and Discussion}
\label{sec:results}

\subsection{Compile Reliability}
\label{sec:res:compile}

HAVEN achieves \textbf{100\% compile success} across all 19 IP designs.
The template-plus-fixup loop produces a VCS-compilable testbench on every run, requiring 1–4 fix iterations (average 1.9).

To confirm that template-generated UVM components are necessary, we conducted an ablation study where all templates were disabled and the LLM generated every UVM component.
None of the three tested designs (ALU, AES, SPI) produced a compiling testbench: the LLM-generated drivers contained blocking/non-blocking misuse, the scoreboard referenced undefined signals, and the monitor sampled on incorrect clock edges.
This confirms that template-generated UVM components are load-bearing for compile success.

\subsection{End-to-End Coverage}
\label{sec:res:e2e}

Table~\ref{tab:coverage} presents HAVEN's code coverage and functional coverage across all 19 IP designs at two pipeline stages: Stage~1 (testbench + predefined sequences) and +Stage~2 (after $K{=}3$ targeted DSL sequence generation iterations).
Functional coverage is measured from the subscriber covergroups generated in Stage~1, parameterized by the Blueprint's \texttt{cover\_bins} declarations, and validated by the three-layer signal consistency check described in Section~\ref{sec:pipe_gen}.

Stage~1's template-driven testbench with predefined sequences reaches 84.6\% average code coverage and 79.8\% average functional coverage on its own.
Stage~2's targeted DSL sequences add +6.0\,pp code and +8.1\,pp functional on average, bringing the final averages to 90.6\% and 87.9\% respectively.

The largest code coverage gains come from designs with the most room to improve after Stage~1: DFI gains +47.6\,pp (41.6\% to 89.2\%) as the LLM composes DDR3-specific initialization sequences, whose multi-phase DDR3 PHY initialization protocol falls outside the predefined sequence patterns, UART gains +24.3\,pp, and UE~SPI gains +16.2\,pp.
Designs showing no Stage~2 gain either already reach high coverage from predefined sequences alone, or require protocol-level state setup that register-level DSL stimuli cannot easily reach (e.g., I2C, HUF).
Per-design breakdowns are in Table~\ref{tab:coverage}; per-protocol averages confirm that the framework generalizes across all three bus protocols without architectural changes.

\subsection{Comparison with Prior Work}
\label{sec:res:ye2025concept}

Table~\ref{tab:landscape} compares HAVEN with the most relevant LLM-based testbench generation systems~\cite{ye2025concept,hu2025uvllm,xu2024meic,qiu2024autobench,qiu2025correctbench,qiu2025confibench}.

\begin{table}[t]
\centering
\caption{Comparison of LLM-based testbench generation systems.}
\label{tab:landscape}
\scriptsize
\setlength{\tabcolsep}{2pt}
\begin{tabular}{@{}l c c c c c@{}}
\toprule
 & \textbf{HAVEN} & \textbf{UVM$^2$} & \textbf{UVLLM} & \textbf{MEIC} & \textbf{AB}$^\dagger$ \\
\midrule
TB format         & UVM     & UVM     & UVM     & Non-UVM & Verilog \\
Template comp.    & All$^\ddagger$ & 5/12  & 0/12    & 0       & 0       \\
Seq.\ mechanism   & DSL     & LLM SV  & LLM SV  & LLM SV  & LLM     \\
Self-correction   & VCS+fix & Regen   & Lint    & ---     & Self    \\
Scale (LOC)       & 180--11k & 1.6k   & Small   & $<$150  & HDLBits \\
\# designs        & 19      & 9       & 3       & 6       & 156     \\
Bus protocols     & 3       & 1+      & ---     & ---     & ---     \\
\bottomrule
\end{tabular}
\vspace{1pt}

{\scriptsize $^\dagger$AB = AutoBench~\cite{qiu2024autobench}/CorrectBench~\cite{qiu2025correctbench}/ConfiBench~\cite{qiu2025confibench}.\\ $^\ddagger$LLM provides only seq\_item field constraints; all other code is template-generated.} \end{table}

Among the systems above, HAVEN is the only one that templates every UVM component slot, and uses a structured DSL for sequence generation instead of free-form LLM SystemVerilog output.
HAVEN also evaluates on the largest and most diverse benchmark: 19 IP designs from 180 to 11\,k~LOC spanning three bus protocols, whereas UVLLM (3 designs), MEIC ($<$150\,LOC per design), and AutoBench/CorrectBench/ConfiBench (HDLBits snippets) are limited to small-scale targets with no standardized bus protocol evaluation.

UVM$^2$~\cite{ye2025concept} is the most closely related system to HAVEN and reports 87.44\% average code coverage and 89.58\% average functional coverage on nine designs.
On the same nine designs, HAVEN achieves 91.0\% code coverage and 90.7\% functional coverage, improving code coverage by 3.6\,pp and functional coverage by 1.1\,pp.

\subsection{Cost and Token Usage}
\label{sec:res:efficiency}

The rightmost columns of Table~\ref{tab:coverage} report per-design LLM call count and API cost for the combined Stage~1 + Stage~2 pipeline on GPT-5.2 (\$1.75 per 1M input tokens, \$14 per 1M output tokens).
Across all 19 IP designs, HAVEN averages 6 LLM calls and \$0.37 per complete testbench, for a total of \$7.08.
Per-design cost ranges from \$0.19 (Simple~SPI) to \$0.62 (CAN), reflecting variation in design specification size and the number of compile-fix and DSL refinement iterations required.
Template rendering itself is free: it accounts for all UVM components and contributes zero LLM tokens.

\subsection{LLM Robustness}
\label{sec:res:llm}

Because HAVEN restricts the LLM to structured information extraction, the framework should not depend on a specific model.
To test this, we re-ran the pipeline with five open-source models (Table~\ref{tab:llm_compare}).
The strongest open-source model, Qwen3.5-27B, reaches 85.8\% code and 78.3\% functional coverage, 4.6\,pp and 7.4\,pp below GPT-5.2; the remaining open-source models land within 79.9--87.3\% code coverage, with the gap concentrated on designs that push LLM output budgets or demand heavy protocol reasoning.
Two bottlenecks dominate: context length and model capacity.
The context-length bottleneck shows up on ETHMAC, which consistently fails at the 16\,K output-token limit on both gpt-oss models during Stage~2 sequence generation.
The capacity bottleneck shows up on the three designs added for the UVM$^2$ comparison (SM4, HUF, DFI): DFI in particular stresses every open model except Qwen3.5-27B, causing Qwen3.5-35B-A3B, Qwen3.5-9B, and gpt-oss-120b to either fail the compile-fix loop or land below 51\% code coverage.
At the smallest scale, Qwen3.5-9B fails to produce a compiling testbench on eight of the nineteen designs because its generated sequence items are too noisy for the compile-fix loop to repair within the iteration budget, and its reported average is over the remaining eleven.
Overall, the framework remains largely model-independent: every open-source model still runs the full HAVEN pipeline and exceeds 79\% code coverage on the designs it handles, but the coverage gap to GPT-5.2 is more pronounced on complex peripherals and the new UVM$^2$-overlap designs.

%% file: sections/7_discussion.tex
\section{Discussion}
\label{sec:discussion}

\textbf{Remaining coverage gaps.}
Six designs remain below 90\% code coverage.
DSL expressiveness limits affect SDRAM (84.0\%), ETHMAC (81.8\%), and HUF (76.5\%), which require multi-phase initialization, DMA descriptor chains, or conditional branching beyond the current linear DSL.
Multi-agent coordination limits affect I2C (75.5\%), whose multi-master arbitration requires coordinated two-agent sequences.
Large state space limits CAN (83.4\%), where cross-domain synchronizer toggles remain undersampled.

\textbf{LLM non-determinism.}
Compile success is deterministic (100\%), but coverage can vary across runs because the sequence item and subscriber are LLM-generated~\cite{zhang2025llm}.
Future work should characterize run-to-run variance.

\textbf{Template engineering effort.}
Each protocol-specific template is a one-time development effort.
In practice, we developed all templates in about one hour using LLM-assisted coding tools with manual verification of protocol-timing correctness.
Once validated, the same template serves all designs under that protocol without per-design modification --- e.g., the single Wishbone driver template covers all eight Wishbone designs from 352 to 11\,k LOC.
All other UVM components (monitor, scoreboard, subscriber) are protocol-agnostic and fully reused across all three protocols.
Adding support for a new bus protocol requires only authoring and validating one new driver template and one new monitor template.\footnote{All templates are available in the \texttt{src/haven/templates/} directory of the released repository.}

\textbf{Threats to validity.}
Our benchmarks are open-source designs that may not reflect proprietary IP complexity~\cite{jin2025realbench,wan2026fixme,liu2023verilogeval,lu2024rtllm}.
Section~\ref{sec:res:llm} shows that HAVEN generalizes across five LLMs with only a 1--8\,pp coverage gap.
The DSL's contribution is quantified by the Stage~1 vs.\ +\,Stage~2 columns in Table~\ref{tab:coverage} (84.6\% $\to$ 90.6\% code, 79.8\% $\to$ 87.9\% functional), providing a built-in ablation.

\section{Conclusion and Future Work}
\label{sec:conclusion}

HAVEN is a hybrid UVM testbench generation framework that delegates all SystemVerilog code generation to rule-based systems while leveraging LLMs solely for structured information extraction.
Generating syntactically correct testbenches using LLM agents is error-prone. To address this, HAVEN employs protocol-specific Jinja2 templates to generate the entire UVM component stack and a Protocol-Aware DSL with rule-based CodeGen to produce correct UVM sequences, significantly reducing LLM calls and token usage.
HAVEN provides state-of-the-art results for LLM-assisted testbench generation. 
Our experiments show HAVEN achieves 100\% compile success, 90.6\% code coverage, and 87.9\% functional coverage across 19 IP designs spanning three sophisticated bus protocols, with an average of only 6 LLM calls and \$0.38 per design; future work will extend the DSL with conditional branching for multi-phase protocols and use LLMs to synthesize new driver templates from protocol specifications.

\section*{Acknowledgements}
The authors would like to thank the Winbond-NYCU Research Center and NTCJ for their support of this work.



%% file: references.bib
@inproceedings{liu2023verilogeval,
  title={Verilogeval: Evaluating large language models for verilog code generation},
  author={Liu, Mingjie and Pinckney, Nathaniel and Khailany, Brucek and Ren, Haoxing},
  booktitle={2023 IEEE/ACM International Conference on Computer Aided Design (ICCAD)},
  pages={1--8},
  year={2023},
  organization={IEEE}
}

@inproceedings{tsai2024rtlfixer,
  title={Rtlfixer: Automatically fixing rtl syntax errors with large language model},
  author={Tsai, YunDa and Liu, Mingjie and Ren, Haoxing},
  booktitle={Proceedings of the 61st ACM/IEEE Design Automation Conference},
  pages={1--6},
  year={2024}
}

@inproceedings{lu2024rtllm,
  title={Rtllm: An open-source benchmark for design rtl generation with large language model},
  author={Lu, Yao and Liu, Shang and Zhang, Qijun and Xie, Zhiyao},
  booktitle={2024 29th Asia and South Pacific Design Automation Conference (ASP-DAC)},
  pages={722--727},
  year={2024},
  organization={IEEE}
}

@article{foster20252024,
  title={2024 Wilson Research Group IC/ASIC functional verification trend report},
  author={Foster, HD},
  journal={Siemens Digital Industries Software},
  year={2025}
}

@article{foster20222022,
  title={The 2022 wilson research group functional verification study},
  author={Foster, Harry and others},
  journal={Siemens. com},
  year={2022}
}

@article{semiconductor2009international,
  title={International technology roadmap for semiconductors},
  author={Semiconductor Industry Association and others},
  journal={http://www. itrs. net},
  year={2009},
  publisher={Semiconductor Industry Association}
}

@article{zhong2023llm4eda,
  title={Llm4eda: Emerging progress in large language models for electronic design automation},
  author={Zhong, Ruizhe and Du, Xingbo and Kai, Shixiong and Tang, Zhentao and Xu, Siyuan and Zhen, Hui-Ling and Hao, Jianye and Xu, Qiang and Yuan, Mingxuan and Yan, Junchi},
  journal={arXiv preprint arXiv:2401.12224},
  year={2023}
}

@article{pan2025survey,
  title={A survey of research in large language models for electronic design automation},
  author={Pan, Jingyu and Zhou, Guanglei and Chang, Chen-Chia and Jacobson, Isaac and Hu, Jiang and Chen, Yiran},
  journal={ACM Transactions on Design Automation of Electronic Systems},
  volume={30},
  number={3},
  pages={1--21},
  year={2025},
  publisher={ACM New York, NY}
}

@inproceedings{xu2024meic,
  title={Meic: Re-thinking rtl debug automation using llms},
  author={Xu, Ke and Sun, Jialin and Hu, Yuchen and Fang, Xinwei and Shan, Weiwei and Wang, Xi and Jiang, Zhe},
  booktitle={Proceedings of the 43rd IEEE/ACM International Conference on Computer-Aided Design},
  pages={1--9},
  year={2024}
}

@inproceedings{hu2025uvllm,
  title={Uvllm: An automated universal rtl verification framework using llms},
  author={Hu, Yuchen and Ye, Junhao and Xu, Ke and Sun, Jialin and Zhang, Shiyue and Jiao, Xinyao and Pan, Dingrong and Zhou, Jie and Wang, Ning and Shan, Weiwei and others},
  booktitle={2025 62nd ACM/IEEE Design Automation Conference (DAC)},
  pages={1--7},
  year={2025},
  organization={IEEE}
}

@inproceedings{zhang2025llm4dv,
  title={Llm4dv: Using large language models for hardware test stimuli generation},
  author={Zhang, Zixi and Szekely, Balint and Gimenes, Pedro and Chadwick, Greg and McNally, Hugo and Cheng, Jianyi and Mullins, Robert and Zhao, Yiren},
  booktitle={2025 IEEE 33rd Annual International Symposium on Field-Programmable Custom Computing Machines (FCCM)},
  pages={133--137},
  year={2025},
  organization={IEEE}
}

@article{thakur2024verigen,
  title={Verigen: A large language model for verilog code generation},
  author={Thakur, Shailja and Ahmad, Baleegh and Pearce, Hammond and Tan, Benjamin and Dolan-Gavitt, Brendan and Karri, Ramesh and Garg, Siddharth},
  journal={ACM Transactions on Design Automation of Electronic Systems},
  volume={29},
  number={3},
  pages={1--31},
  year={2024},
  publisher={ACM New York, NY}
}

@article{liu2023chipnemo,
  title={Chipnemo: Domain-adapted llms for chip design},
  author={Liu, Mingjie and Ene, Teodor-Dumitru and Kirby, Robert and Cheng, Chris and Pinckney, Nathaniel and Liang, Rongjian and Alben, Jonah and Anand, Himyanshu and Banerjee, Sanmitra and Bayraktaroglu, Ismet and others},
  journal={arXiv preprint arXiv:2311.00176},
  year={2023}
}

@inproceedings{yan2025assertllm,
  title={Assertllm: Generating hardware verification assertions from design specifications via multi-llms},
  author={Yan, Zhiyuan and Fang, Wenji and Li, Mengming and Li, Min and Liu, Shang and Xie, Zhiyao and Zhang, Hongce},
  booktitle={Proceedings of the 30th Asia and South Pacific Design Automation Conference},
  pages={614--621},
  year={2025}
}

@inproceedings{xu2025revolution,
  title={Revolution or Hype? Seeking the Limits of Large Models in Hardware Design},
  author={Xu, Qiang and Stok, Leon and Drechsler, Rolf and Wang, Xi and Zhang, Grace Li and Markov, Igor L},
  booktitle={2025 IEEE/ACM International Conference On Computer Aided Design (ICCAD)},
  pages={1--9},
  year={2025},
  organization={IEEE}
}

@article{zhang2025llm,
  title={Llm hallucinations in practical code generation: Phenomena, mechanism, and mitigation},
  author={Zhang, Ziyao and Wang, Chong and Wang, Yanlin and Shi, Ensheng and Ma, Yuchi and Zhong, Wanjun and Chen, Jiachi and Mao, Mingzhi and Zheng, Zibin},
  journal={Proceedings of the ACM on Software Engineering},
  volume={2},
  number={ISSTA},
  pages={481--503},
  year={2025},
  publisher={ACM New York, NY, USA}
}

@article{zhang2026understanding,
  title={Understanding and Mitigating Errors of LLM-Generated RTL Code},
  author={Zhang, Jiazheng and Liu, Cheng and Cheng, Long and Li, Xiaowei and Li, Huawei},
  journal={IEEE Transactions on Computer-Aided Design of Integrated Circuits and Systems},
  year={2026},
  publisher={IEEE}
}

@article{tonmoy2024comprehensive,
  title={A comprehensive survey of hallucination mitigation techniques in large language models},
  author={Tonmoy, SMTI and Zaman, SM and Jain, Vinija and Rani, Anku and Rawte, Vipula and Chadha, Aman and Das, Amitava},
  journal={arXiv preprint arXiv:2401.01313},
  volume={6},
  year={2024}
}

@inproceedings{qiu2024autobench,
  title={Autobench: Automatic testbench generation and evaluation using llms for hdl design},
  author={Qiu, Ruidi and Zhang, Grace Li and Drechsler, Rolf and Schlichtmann, Ulf and Li, Bing},
  booktitle={Proceedings of the 2024 ACM/IEEE International Symposium on Machine Learning for CAD},
  pages={1--10},
  year={2024}
}

@inproceedings{qiu2025correctbench,
  title={Correctbench: Automatic testbench generation with functional self-correction using llms for hdl design},
  author={Qiu, Ruidi and Zhang, Grace Li and Drechsler, Rolf and Schlichtmann, Ulf and Li, Bing},
  booktitle={2025 Design, Automation \& Test in Europe Conference (DATE)},
  pages={1--7},
  year={2025},
  organization={IEEE}
}

@article{qiu2025confibench,
  title={ConfiBench: Automatic Testbench Generation with Confidence-Based Scenario Mask and Testbench Ensemble using LLMs for HDL Design},
  author={Qiu, Ruidi and Zhang, Grace Li and Drechsler, Rolf and Ho, Tsungyi and Schlichtmann, Ulf and Li, Bing},
  journal={ACM Transactions on Design Automation of Electronic Systems},
  year={2025},
  publisher={ACM New York, NY}
}

@article{yao2025hdldebugger,
  title={Hdldebugger: Streamlining hdl debugging with large language models},
  author={Yao, Xufeng and Li, Haoyang and Chan, Tsz Ho and Xiao, Wenyi and Yuan, Mingxuan and Huang, Yu and Chen, Lei and Yu, Bei},
  journal={ACM Transactions on Design Automation of Electronic Systems},
  volume={30},
  number={6},
  pages={1--26},
  year={2025},
  publisher={ACM New York, NY}
}

@article{ieee1800ieee,
  title={IEEE standard for universal verification methodology language reference manual},
  author={IEEE Standards Association and others},
  journal={IEEE Std},
  volume={2020},
  pages={1--458},
  year={1800}
}

@article{ye2025concept,
  title={From Concept to Practice: an Automated LLM-aided UVM Machine for RTL Verification},
  author={Ye, Junhao and Hu, Yuchen and Xu, Ke and Pan, Dingrong and Chen, Qichun and Zhou, Jie and Zhao, Shuai and Fang, Xinwei and Wang, Xi and Guan, Nan and others},
  journal={arXiv preprint arXiv:2504.19959},
  year={2025}
}

@inproceedings{gadde2024efficient,
  title={Efficient stimuli generation using reinforcement learning in design verification},
  author={Gadde, Deepak Narayan and Nalapat, Thomas and Kumar, Aman and Lettnin, Djones and Kunz, Wolfgang and Simon, Sebastian},
  booktitle={2024 20th International Conference on Synthesis, Modeling, Analysis and Simulation Methods and Applications to Circuit Design (SMACD)},
  pages={1--4},
  year={2024},
  organization={IEEE}
}

@inproceedings{fine2003coverage,
  title={Coverage directed test generation for functional verification using bayesian networks},
  author={Fine, Shai and Ziv, Avi},
  booktitle={Proceedings of the 40th annual Design Automation Conference},
  pages={286--291},
  year={2003}
}

@inproceedings{kumar2023optimizing,
  title={Optimizing constrained random verification with ml and bayesian estimation},
  author={Kumar, Bhuvnesh and Parthasarathy, Ganapathy and Nanda, Saurav and Rajakumar, Sridhar},
  booktitle={2023 ACM/IEEE 5th Workshop on Machine Learning for CAD (MLCAD)},
  pages={1--6},
  year={2023},
  organization={IEEE}
}

@article{wu2024survey,
  title={Survey of machine learning for software-assisted hardware design verification: Past, present, and prospect},
  author={Wu, Nan and Li, Yingjie and Yang, Hang and Chen, Hanqiu and Dai, Steve and Hao, Cong and Yu, Cunxi and Xie, Yuan},
  journal={ACM Transactions on Design Automation of Electronic Systems},
  volume={29},
  number={4},
  pages={1--42},
  year={2024},
  publisher={ACM New York, NY}
}

@article{ioannides2012coverage,
  title={Coverage-directed test generation automated by machine learning--a review},
  author={Ioannides, Charalambos and Eder, Kerstin I},
  journal={ACM Transactions on Design Automation of Electronic Systems (TODAES)},
  volume={17},
  number={1},
  pages={1--21},
  year={2012},
  publisher={ACM New York, NY, USA}
}

@article{wang2023grammar,
  title={Grammar prompting for domain-specific language generation with large language models},
  author={Wang, Bailin and Wang, Zi and Wang, Xuezhi and Cao, Yuan and A Saurous, Rif and Kim, Yoon},
  journal={Advances in Neural Information Processing Systems},
  volume={36},
  pages={65030--65055},
  year={2023}
}

@article{geng2025generating,
  title={Generating structured outputs from language models: Benchmark and studies},
  author={Geng, Saibo and Cooper, Hudson and Moskal, Micha{\l} and Jenkins, Samuel and Berman, Julian and Ranchin, Nathan and West, Robert and Horvitz, Eric and Nori, Harsha},
  journal={arXiv e-prints},
  pages={arXiv--2501},
  year={2025}
}

@article{tasiran2002coverage,
  title={Coverage metrics for functional validation of hardware designs},
  author={Tasiran, Serdar and Keutzer, Kurt},
  journal={IEEE Design \& Test of Computers},
  volume={18},
  number={4},
  pages={36--45},
  year={2002},
  publisher={IEEE}
}

@article{naveh2007constraint,
  title={Constraint-based random stimuli generation for hardware verification},
  author={Naveh, Yehuda and Rimon, Michal and Jaeger, Itai and Katz, Yoav and Vinov, Michael and s Marcu, Eitan and Shurek, Gil},
  journal={AI magazine},
  volume={28},
  number={3},
  pages={13--13},
  year={2007}
}

@article{zhang2026llm4cov,
  title={LLM4Cov: Execution-Aware Agentic Learning for High-coverage Testbench Generation},
  author={Zhang, Hejia and Yu, Zhongming and Ho, Chia-Tung and Ren, Haoxing and Khailany, Brucek and Zhao, Jishen},
  journal={arXiv preprint arXiv:2602.16953},
  year={2026}
}

@article{nadimi2025tb,
  title={TB or Not TB: Coverage-Driven Direct Preference Optimization for Verilog Stimulus Generation},
  author={Nadimi, Bardia and Filom, Khashayar and Chen, Deming and Zheng, Hao},
  journal={arXiv preprint arXiv:2511.15767},
  year={2025}
}

@article{jin2025realbench,
  title={Realbench: Benchmarking verilog generation models with real-world ip designs},
  author={Jin, Pengwei and Huang, Di and Li, Chongxiao and Cheng, Shuyao and Zhao, Yang and Zheng, Xinyao and Zhu, Jiaguo and Xing, Shuyi and Dou, Bohan and Zhang, Rui and others},
  journal={arXiv preprint arXiv:2507.16200},
  year={2025}
}

@inproceedings{wan2026fixme,
  title={Fixme: Towards end-to-end benchmarking of llm-aided design verification},
  author={Wan, Gwok-Waa and Wong, SamZaak and Su, Shengchu and Niu, Chenxu and Wang, Ning and Wan, Xinlai and Chen, Qixiang and Xing, Mengnv and Zhang, Jingyi and Ye, Jianmin and others},
  booktitle={Proceedings of the AAAI Conference on Artificial Intelligence},
  volume={40},
  number={2},
  pages={1087--1095},
  year={2026}
}

@article{chang2023chipgpt,
  title={Chipgpt: How far are we from natural language hardware design},
  author={Chang, Kaiyan and Wang, Ying and Ren, Haimeng and Wang, Mengdi and Liang, Shengwen and Han, Yinhe and Li, Huawei and Li, Xiaowei},
  journal={arXiv preprint arXiv:2305.14019},
  year={2023}
}
